\documentclass[12pt]{article}
\usepackage[english]{babel}
\usepackage{mathtext}
\usepackage[T2A]{fontenc}
\begin{document}
\begin{center} {\Large {\bf Maximum entropy principle for Renyi's and Tsallis' entropies}}\\
  \vspace{.2cm}
\renewcommand{\thefootnote}{\fnsymbol{footnote}}
\vspace{.5cm} \large {\bf A.G.Bashkirov}\footnote{{\it E-mail
address}: abas@idg.chph.ras.ru}\\ \vspace{.25cm} Institute
Dynamics of Geospheres, RAS,\\ Leninskii prosp. 38 (bldg.1),
119334, Moscow, Russia\\
\end{center}
\begin{abstract}
The equilibrium distributions of probabilities providing
maximality of Renyi and Tsallis entropies are rederived. New
S-forms of them are found which are normalised with corresponding
entropies in contrast to the usual Z-forms normalised with
partition functions.
\end{abstract}
 \vspace{.1cm}
 \setcounter{footnote}{0}

\section{Introduction}

According to the well-known maximum entropy principle (MEP)
developed by Jaynes [1] for a Boltzmann-Gibbs statistics the
distribution of probabilities $p=\{p_i\}$ provides maximality of
information entropy with an additional conditions which consist in
preassigning the average value $ U=\langle H\rangle_p\equiv\sum_i
H_i p _i$ of the random quantity $H_i$ and the requirement of
normalisation of $p$.

Then, the distribution $\{p_i\}$ is determined from the extremum
of the functional
 \begin{equation}
L_G(p )=- \sum_i^W\,p_i\ln p_i - \beta_0  \,\sum_i^W H_i p _i -
\alpha_0 \sum_i^W\,p _i;
\end{equation}
where $\beta_0 $ and $ \alpha_0 $ are Lagrange multipliers. Its
extremum is ensured by the Gibbs canonical distribution, in which
$\beta_0 =1/k_BT_0$ and $\alpha_0+1 $ is the free energy.

Really, equating its functional derivative to zero we get
\begin{equation}
\frac{\delta L_G(p )}{\delta p _i}=-\ln p_i-\beta_0 H_i - \alpha_0
-1 =0.
\end{equation}
or
\begin{equation}
p_i=e^{-\alpha_0-1- \beta_0 H_i}
\end{equation}
With account of normalization condition $\sum_i p_i=1$ we get
\begin{equation}
e^{\alpha_0+1}=Z_G\equiv \sum_i e^{- \beta_0 H_i}
\end{equation}
and
\begin{equation}
p_i=Z_G^{-1}e^{- \beta_0 H_i}
\end{equation}
On the other side, multiplying equation (2) by $p_i$ and summing
up over $i$, with account of normalization condition $\sum_i
p_i=1$ we get
\begin{equation}
\alpha_0+1=S_B/k_B- \beta_0  U
\end{equation}
and
\begin{equation}
p_i=e^{-S_B/k_B+ \beta_0  U- \beta_0 H_i}
\end{equation}
When $\beta_0=1/k_BT_0$ both equations (5) and (7) are equivalent
due to definitions of a free energy $F= U - T_0S$ and
$F=-k_BT_0\ln Z_G$. The first of them, eq. (5), may be called a
Z-form and the second one, eq. (7), a S-form of the Gibbs'
distribution.

\section{MEP for Renyi entropy}

If the Renyi entropy $S_R=\frac {k_B}{1-q}\ln \sum_i p_i^{q }$ is
used instead of the Boltzmann entropy the equilibrium distribution
must provide maximum of the functional
\begin{equation}
L_R(p )=\frac 1{1-q}\,\ln \sum_i^W\,p^{q } _i - \beta \,\sum_i^W
H_i p _i - \alpha \sum_i^W\,p _i,
\end{equation}
where $\beta $ and $ \alpha $ are Lagrange multipliers. Note that,
in the $q\to 1$ limit $L_R(p)$ passes to $L_R(p )$.

We equate a functional derivative of $L_R(p )$ to zero, then
\begin{equation}
\frac{\delta L_R(p )}{\delta p _i}=\frac q{1-q}\,\frac
{p_i^{q-1}}{\sum_j p_j^{q}}-\beta H_i - \alpha =0.
\end{equation}
Multiplying this equation by $p_i$ and summing up over $i$, with
account of normalisation condition $\sum_i p_i=1$ we get $\alpha
=\frac q{1-q}- \beta   U$. Then, it follows from equation (9) that
\begin{equation}
p _i=\left(\sum_j^W\,p^{q} _j\, (1-\beta\frac{q-1}{q} \Delta H_i)
\right)^{\frac{1}{q-1}},\,\,\Delta H_i=H_i-  U.
\end{equation}
Using once more the condition $\sum_i p_i=1$ we get $$
\sum_j^W\,p_j^q =\left(\sum_i^W (1-\beta\frac{q-1}{q}\Delta
H_i)^{\frac{1}{q-1}}\right)^{-(q-1)}$$ and, finally,
\begin{eqnarray}
p_i=p_i^{(RZ)} &=&Z_R^{-1}\left(1-\beta\frac{q-1}{q}\Delta
H_i\right)^{\frac{1}{q-1}}\\
Z_R^{-1}&=&\sum_i\left(1-\beta\frac{q-1}{q}\Delta
H_i\right)^{\frac{1}{q-1}}\nonumber
\end{eqnarray}
This is the Renyi distribution in the Z-form. At $q\to 1$ the
distribution $\{p^{(R)}_i\}$ becomes the Z-form of Gibbs canonical
distribution in which the constant $\beta=1/k_BT_0$ is the
reciprocal of the temperature.

To obtain the S-form of the Renyi distribution we can find that
for the Renyi distribution (11) $S_R=\ln Z_R$ or rewrite the sum
$\sum_jp^{q}_j$ as $\sum_jp^{q}_j=\exp\{(1-q)S_R/k_B\}$. Then we
get
\begin{equation}
p^{(RS)}_i=e^{-S_R/k_B}\left(1-\beta\frac{q-1}{q} \Delta H_i
\right)^{\frac{1}{q-1}}.
\end{equation}
At $q\to 1$ this distribution becomes the S-form of Gibbs
canonical distribution.

\section{MEP for Tsallis entropy}

When the Tsallis entropy was used firstly instead of the Boltzmann
entropy the equilibrium distribution was derived by Tsallis [2] in
the form
\begin{equation}
p^{T}_i =\frac {\left(1+\beta(1-q)H_i\right)^{\frac{1}{q-1}}}
{\sum_i\left(1+\beta(1-q)H_i\right)^{\frac{1}{q-1}}},
\end{equation}
This distribution was named later as the 1-st version of
thermostatistics. It was noticed there that the parameter $\beta$
was not a Lagrange multiplier because of the starting functional
was forced to be taken as
\begin{equation} L^{T}(p )=-\frac
1{1-q}\,\left(1- \sum_i^W\,p^{q } _i\right) - \alpha\beta (q-1)
\,\sum_i^W H_i p _i + \alpha \sum_i^W\,p _i.
\end{equation}

Here, the question arises about forms of Lagrange multipliers
$\alpha\beta(q-1)$ and $(+\alpha)$, but the main problem is that
the functional $L^{T}(p )$ does not pass to the functional (9)
when $q\to 1$ as the second term in (13) vanishes.

It seems reasonable to suppose that just this difficulty caused to
introduce the 3-rd version of nonextensive thermodynamics with the
escort distribution  $P_i=p_i^q/\sum_ip_i^q$ . The consistency of
the transition to the escort distribution is partly justified by
the condition conservation of a preassigned average value of the
energy $ U=\langle H\rangle_{es}\equiv\sum_i H_i\,P_i$, however
other average values are to be calculated with the use of the same
escort distribution also, that contradicts to the main principles
of probability description.

For the generality sake, the resulted distribution of the Tsallis'
3-rd version is reproduced here as
\begin{eqnarray}
p_i^{T3}& =&Z_{T3}^{-1}\left(1-\beta (1-q)(H_i- U
)/\sum_j\,(p_j^{T3})^{q} \right)^{\frac{1}{1-q}},\\
Z_{T3}&=&\sum_i\left(1-\beta (1-q)(H_i- U
)/\sum_j\,(p_j^{T3})^{q}\right)^{\frac{1}{1-q}}
\end{eqnarray}
where $\beta$ is the true Lagrange multiplier in the corresponding
variational functional of the 3-rd version.

Below, it will be shown that careful solution of the variational
problem for the 1-st version of thermostatistics gives rise to a
probability distribution which doesn't seem to be less acceptable
then $p_i^{T3}$.

Really, if we take the functional
\begin{equation}
L_T(p )=\frac 1{1-q}(\sum_i^W\,p^{q } _i -1) - \beta \,\sum_i^W
H_i p _i - \alpha \sum_i^W\,p _i,
\end{equation}
and equate its functional derivative to zero we get
\begin{equation}
\frac{\delta L_T(p )}{\delta p _i}=\frac q{1-q} p_i^{q-1}-\beta
H_i - \alpha =0.
\end{equation}
Multiplying this equation by $p_i$ and summing up over $i$, with
account of normalisation condition $\sum_i p_i=1$ we get $\alpha
=\frac q{1-q}\sum_i\,p_i^{q }- \beta   U$. Then, it follows from
equation (18) that
\begin{eqnarray}
p _i &=&\left(\sum_j\,p_j^{q} -\beta\frac{q-1}{q} \Delta H_i
\right)^{\frac{1}{q-1}}\nonumber\\ {}&=&\left(\sum_j\,p_j^{q}
\right)^{\frac{1}{q-1}}\, \left(1-\beta\frac{q-1}{q} \Delta
H_i/\sum_j\,p_j^{q}) \right)^{\frac{1}{q-1}}
\end{eqnarray}
Using once more the condition $\sum_i p_i=1$ we get
\begin{eqnarray}
p_i=p_i^{TZ} &=&Z_{T1}^{-1}\left(1-\beta\frac{q-1}{q}\Delta
H_i/\sum_j\,p_j^{q}\right)^{\frac{1}{q-1}},\\
Z_{T1}&=&\sum_i\left(1-\beta\frac{q-1}{q}\Delta
H_i/\sum_j\,p_j^{q}\right)^{\frac{1}{q-1}}
\end{eqnarray}
This is my modification of the 1-st version of the Tsallis
distribution in the Z-form. It differs from the 3-rd version, eq.
(15), by the signs before differences $q-1$ and doesn't invoke the
escort distribution. In this respect it seems to be much more
attractive then $p^{T3}$. At $q\to 1$ the distribution $p^{TZ}$
becomes the Z-form of the Gibbs canonical distribution.

It should be noted that both $p^{T3}$ and $p^{TZ}$ are explicitly
self-referential in contrast to $p^{RS}$.

To obtain the S-form of the modified 1-st version of the Tsallis
distribution we rewrite the sum $\sum_j\,p_j^{q}$ in the upper
line of eq. (19) in term of the Tsallis entropy. Then we get
finally
\begin{equation}
p_i=p _i^{TS}=\left(1-(q-1)(\frac {S_T}{k_B}-\frac {\beta}{q} U
+\frac {\beta}{q}H_i) \right)^{\frac{1}{q-1}}.
\end{equation}
At $q\to 1$ this distribution becomes the S-form of the Gibbs
canonical distribution.

\section{Conclusion}
In this article, I derived and rederived some useful equations for
equilibrium probability distributions on the base of MEP for the
Renyi and Tsallis entropies. I hope that they can abandon the use
of escort distributions for calculations of average values in
favour of original distributions.

Besides, the new S-forms of the Renyi and Tsallis distributions
are proposed. They are normalised with the use of corresponding
entropies but not partition functions. Theese forms may appear to
be useful in different applications and in construction of new,
non-Gibbsian, thermostatistics.

\subsection*{Acknowledgements} It is pleasure to thank A. Vityazev
for fruitful discussions and supporting this work, and I
acknowledge warm, friendly and illuminating discussions with S.
Abe, C. Beck, A. Wang and G. Wilk held at NEXT2003 on Sardgenia.

\subsection*{References}
[1] Jaynes E.T. Phys.Rev. {\bf 106}, 620; {\bf 107}, 171
(1957)\\[0pt] [2] Tsallis C. J.Stat.Phys. {\bf 52}, 479
(1988)\\[0pt]
\end{document}